# Nitrogen and hydrogen intercalation into crystalline fullerite $C_{60}$ and photoluminescent studies in a wide temperature range

V. Zoryansky, P. Zinoviev and Yu. Semerenko.

*B. Verkin Institute for Low Temperature Physics and Engineering of the National Academy of Sciences of Ukraine, Kharkiv 61103, Ukraine*

zoryansky@gmail.com

**Abstract:** The optical properties of $C_{60}$ single crystals saturated with hydrogen and nitrogen molecules were studied by the spectral-luminescence method in the temperature range of 20–230 K. The saturation was carried out under a pressure of 30 atm at different temperatures from 200 °C to 450 °C. Early it have been established of $C_{60}$ fullerite intercalated with $H_2$ and $N_2$ molecules that the temperature limit of the adsorption crossover is about 250 °C and 420 °C, respectively (transition from the diffusion mechanism of intercalation - physisorption to chemical interaction - chemisorption). At saturation temperatures higher than this temperature limit the process of chemical interaction of $H_2$ or $N_2$ impurity molecules and the $C_{60}$ matrix occurs with the formation of a new chemical compounds - $C_{60}H_x$ and $(C_{59}N)_2$. For the first time we present the results of the photoluminescent properties of new $C_{60}$ fullerite intercalated with $N_2$ and $H_2$ molecules - hydrofullerite $C_{60}H_x$ and biazafullerite $(C_{59}N)_2$. In $C_{60}H_x$ compound the integrated radiation intensity is independent from temperature has been recorded. This behavior of integrated radiation intensity have been explained by the absence of an orientational phase transition and the transition to a glassy state in hydrofullerite. In $(C_{59}N)_2$ compound the temperature dependence of the integral radiation intensity has been registered. Also in biazafullerite the low temperature quenching of photoluminescence has been detected. The new effect of low-temperature quenching of photoluminescence has been explained by the appearance of effective capture centers of the excitons which occurs as a result of the chemical interaction of $N_2$ impurity molecules and the $C_{60}$ matrix, and the non-radiative deactivation of electronic excitation.

# 1. Introduction

Fullerite $C_{60}$ belongs to the class of molecular crystals. Due to the peculiarities of the crystal structure, molecular crystals, both simple and complex, often have adsorption [1], polymerization [2], nonlinear optical and other unique physical properties. In particular, fullerite $C_{60}$ has an unusually high ability to adsorb simple gases. Unlike classical molecular crystals (where the molecules are quite rigidly fixed in the crystal lattice), $C_{60}$ molecules in fullerite crystals can rotate. The mutual orientational arrangement of neighboring molecules significantly affects the transfer of electronic excitations in the crystal lattice and depends significantly on the presence of impurities in intermolecular voids. The molecular and crystalline structure of $C_{60}$ determine two main ways of saturating it with simple molecules - filling intermolecular intra-lattice voids (physical sorption) and the formation of new substances as a result of a direct chemical reaction of impurity molecules and fullerene (chemical sorption). Spectral characteristics of fullerite are sensitive both to the degree of filling of intermolecular voids with impurity and to the formation of new chemical bonds. This makes fullerite $C_{60}$ a convenient material for studying lattice dynamics by spectral-luminescent methods under the influence of various intercalants. Both processes of saturation and interaction are accelerated with increasing temperature and pressure. Based on previously conducted studies [3, 4], it is possible to identify temperature ranges in which fullerite saturation with hydrogen and nitrogen occurs mainly due to physical diffusion or chemical interaction. At temperatures T<250 °C, the filling of intermolecular cavities with molecular hydrogen occurs due to the first mechanism. For nitrogen, this temperature limit is 420 °C. At temperatures above these values, the chemical interaction of fullerene molecules with hydrogen and nitrogen, respectively, becomes decisive. These processes inevitably affect the radiative properties of the indicated two-component complexes. As was established earlier [5], for efficient transport of excitons to emission centers in $C_{60}$ crystals, the coherence of states on two neighbouring molecules participating in the exciton tunnel jump is important, which is observed with their certain mutual orientation. This condition is well satisfied in the low-temperature (glassy) phase of pure fullerite, in which the pentagonal orientation of $C_{60}$ molecules predominates, up to the orientational ordering (glass transition) temperature $T_g$. As the temperature increases above $T_g$, the probability of molecule reorientation increases, which inevitably leads to a violation of the coherence of states, a decrease in the exciton mean free path and, as a consequence, to an increase in the probability of excitation deactivation without radiation. It is for this reason that the integrated radiation intensity I in the low-temperature phase of pure $C_{60}$ is maximum and is practically independent of temperature up to $T_g$. Above the glass transition temperature $T_g$, a rather sharp drop in the radiation intensity occurs, progressing with heating [5, 6]. The main contribution to the decrease in I at T>$T_g$ is made by the luminescence of localized states

("deep X-traps"), the concentration of which is determined, among other things, by the degree and nature of filling of intermolecular voids with impurity particles. Our results [5] showed that the diffusion saturation mechanism for $H_2$ differs significantly from that for $N_2$. Hydrogen saturation occurs in two stages, with the octahedral voids being filled with one molecule in the early stage. After this, double filling of these voids begins, leading to additional impurity-matrix interaction. The presence of two particles in one intermolecular void has an additional braking effect on the rotational motion of fullerene molecules in the crystal lattice sites. As a result, the integral luminescence intensities, as a function of temperature $I$(T), begin to change significantly. Unlike pure $C_{60}$, in which the spectra begin to fade at the glass transition point $T_g$ =95 K, for the solid solution $C_{60}+H_2$, with increasing temperature, the intensity remains almost constant up to higher temperatures. The threshold temperature is higher the longer the hydrogen saturation time. Consequently, $T_g$ increases until thermal vibrations of the lattice are able to initiate molecular rotations. In the case of $N_2$, the molecules are too large for two particles to occupy one octahedral cavity. They can only slightly expand the lattice [7-9], thereby facilitating the rotations of the fullerite molecules. This is due to the influence of intracrystalline "negative pressure", which makes it possible to significantly facilitate the reorientation of $C_{60}$ molecules. Therefore, as our experiments on the luminescence of the $C_{60}+N_2$ complex show, the $T_g$ point shifts toward lower temperatures, which, unlike the case of hydrogen saturation, indicates a weak impurity-matrix interaction with a simultaneous weakening of the bond between $C_{60}$ molecules in the crystal lattice. With an increase in the saturation temperature, the diffusion mechanism of sorption into the fullerite lattice for both hydrogen and nitrogen undergoes a transition to a chemical interaction of the impurity with the matrix (chemisorption) and an adsorption crossover is observed.

This article presents the results of studies of the temperature behaviour of the photoluminescent properties of $C_{60}$ fullerite crystals doped with $H_2$ and $N_2$, upon transition from the diffusion saturation mechanism to the conditions of chemical sorption from the gas phase. For the first time, the temperature dependences of the integral radiation intensity of a new hydrogen-containing chemical compound $C_{60}H_x$ have been established. Based on luminescent data, a comparative analysis of the mechanisms of interaction of hydrogen and nitrogen impurities with the $C_{60}$ matrix in two saturation modes was carried out.

**2. Methods and materials**

For the experiments on studying the temperature behaviour of the photoluminescence of the $C_{60}+H_2$ and $C_{60}+N_2$ complexes, polycrystalline samples in the form of powder with a granule size of about 0.5 mm and a purity of at least 99.9% were used. To remove residual atmospheric gases before saturation, the fullerite was preliminarily kept in a dynamic vacuum of $10^{-3}$ mm Hg at a temperature of 300 °C for 48 h. After such annealing before saturation with $H_2$ and $N_2$, photoluminescence spectra were recorded that corresponded to the spectra of pure crystalline fullerite $C_{60}$ [6] with virtually no impurities. This allowed us to conclude that the samples were sufficiently pure. Then, immediately before saturation with hydrogen and nitrogen, the samples were again subjected to additional degassing for four hours at a temperature of 300 °C in a dynamic vacuum. In accordance with the X-ray structural analysis data on the change in the lattice parameter upon saturation of $C_{60}$ with $N_2$ molecules [3, 4], intercalation modes were selected that clearly corresponded to the mechanisms of either only physical sorption for both impurities or those that already ensured a confident impurity-matrix chemical interaction. Thus, the samples were saturated with nitrogen at two different temperatures 280 °C and 450 °C. The total time of powder intercalation under gas pressure in the cell P=30 atm. at each of these temperatures was limited by the moment of significant slowdown in the growth of the lattice parameter, which corresponds to impurity concentrations close to the maximum possible for the specified modes [4]. The selected intercalation conditions correspond to the temperature regions of saturation with a slow and rapid increase in the lattice parameter of fullerite $C_{60}$, respectively below and above the adsorption crossover temperature at 420 °C. As shown earlier, the characteristic break in the curve of the dependence of the lattice parameter change on the saturation temperature at a fixed pressure near 420 °C indicates a change in the sorption mechanism. A detailed description of the process of saturation of $C_{60}$ with $N_2$ molecules and the data of X-ray structural analysis of the obtained samples are given in [4]. For the samples of fullerite $C_{60}$ saturated with hydrogen $H_2$, modes with the same conditions of unambiguous correspondence to either physical or chemical sorption were selected. Thus, according to [3], a purely diffusion mechanism of intercalation at a gas pressure in the cell of P=30 atm corresponds to a mode with a system temperature of 200 °C. From the totality of the obtained results it follows that at P=30 atm in the region T≤250 ºC the probability of chemical interaction of hydrogen with $C_{60}$ is extremely small. In this temperature range, only diffusion filling of intermolecular cavities with $H_2$ molecules and the formation of a solid interstitial solution occur. Also, the previously obtained results on $C_{60}$ single crystals [10] allow us to consider that 250 °C is the boundary temperature above which the mechanism of chemical sorption of hydrogen by fullerite is activated and increases with increasing temperature. Accordingly, to obtain a stable chemical interaction of $C_{60}$ and $H_2$, the temperature was increased to 300 °C at the same pressure P=30 atm. Photoluminescence of $C_{60}$ crystals was recorded "in reflection" in the spectral region of 1.2–1.85 eV (1033–670 nm) with a spectral resolution of 2 nm using a high-aperture diffraction monochromator MDR-2 with an electromechanical drive and a cooled photomultiplier PEM-62 (spectral characteristic type S-1) in the photon counting mode. The cryogenic part of the experimental setup allowed changing the sample temperature in a wide temperature range of 20–230 K and stabilizing it during the experiment with an accuracy of 0.5 K. A He-Ne laser with $E_{exc}$ = 1.96 eV (632.8 nm) was used to excite photoluminescence. The excitation power density was W ≤ 1 mW/mm$^2$. Such a limitation of the excitation power was introduced to prevent undesirable photostimulated processes in the surface layers of the studied polycrystalline samples. The luminescence measurement technique, the experimental setup, and the analytical processing of the experimental results are presented in [5].

## 3. Results and discussion

As mentioned above, the luminescence of fullerite at low temperatures is determined by the efficiency of exciton transport to the emission centers, usually to structural defects and impurities, which have the general name of "deep X-traps" [6]. Localization of the Frenkel exciton on such inclusions sharply increases the probability of radiative transitions forbidden by symmetry for $C_{60}$ molecules in an ordered lattice [5]. In the case of significant filling of the intermolecular voids of the fullerite lattice with neutral impurities, such as hydrogen or nitrogen, the efficiency of the emission centers will most likely increase. In the article [3], it is shown that for almost all sorption modes with saturation temperatures below 250 ºC, for which the holding of samples in hydrogen under the same pressure P=30 atm. lasted until the new lattice parameter stabilized, almost the same lattice deformation values were obtained, where Δa reached ~ 0.058 Å or Δa/a0 = 0.4%. This indicates that all octahedral cavities in $C_{60}$ crystallites were filled with molecular hydrogen and an equilibrium, highly saturated interstitial solution was formed. For the case of nitrogen, the process of diffusion saturation differs significantly from that with hydrogen, mainly due to the physical dimensions of the intercalant molecule. With an increase in the saturation temperature of $C_{60}$ fullerite with nitrogen molecules up to 420 °C, the octahedral voids of the cubic lattice are gradually filled with a large density gradient of the dopant along the depth of the sample [11]. Taking into account the fact that the exciting radiation penetrates only into the surface layers of the crystals under study, we can confidently assume that in both cases, for hydrogen and nitrogen, we recorded the spectra of saturated solid solutions with a concentration of radiating centers close to the maximum. This will inevitably be reflected in the photoluminescence spectra of the $C_{60}+H_2$ and $C_{60}+N_2$ complexes. At sorption temperatures above 250 °C for hydrogen and 420 °C for nitrogen, respectively, the diffusion saturation mechanism is most likely replaced by direct chemical interaction between the molecules of the matrix and the impurity. The formation of hydrogen-containing and nitrogen-containing new substances [4] occurs, with photoluminescent characteristics completely different from those of the intercalation solutions.

## 3.1 $C_{60}+H_2$ system.

Figure 1 shows the photoluminescence spectra of single crystals of pure fullerite $C_{60}$ saturated with hydrogen in the modes of physical - 200 °C and chemical - 300 °C sorption, under a pressure of 30 atm. It was found that for the low-temperature spectrum (T=10 K) of the solid solution, the following effects are observed: a noticeable change and redistribution of the intensity of the spectral bands, their inhomogeneous broadening and a shift of the entire photoluminescence spectrum to the region of low energies occurs; annealing in a vacuum leads to an almost complete restoration of the initial form of the photoluminescence spectra [12], characteristic of pure fullerite. In the case of saturation of fullerite $C_{60}$ with hydrogen at 300 °C, the photoluminescence spectrum at 10 K has a diffuse form and shifts to the region of higher energies. The latter indicates an increase in intermolecular interaction in the crystals.

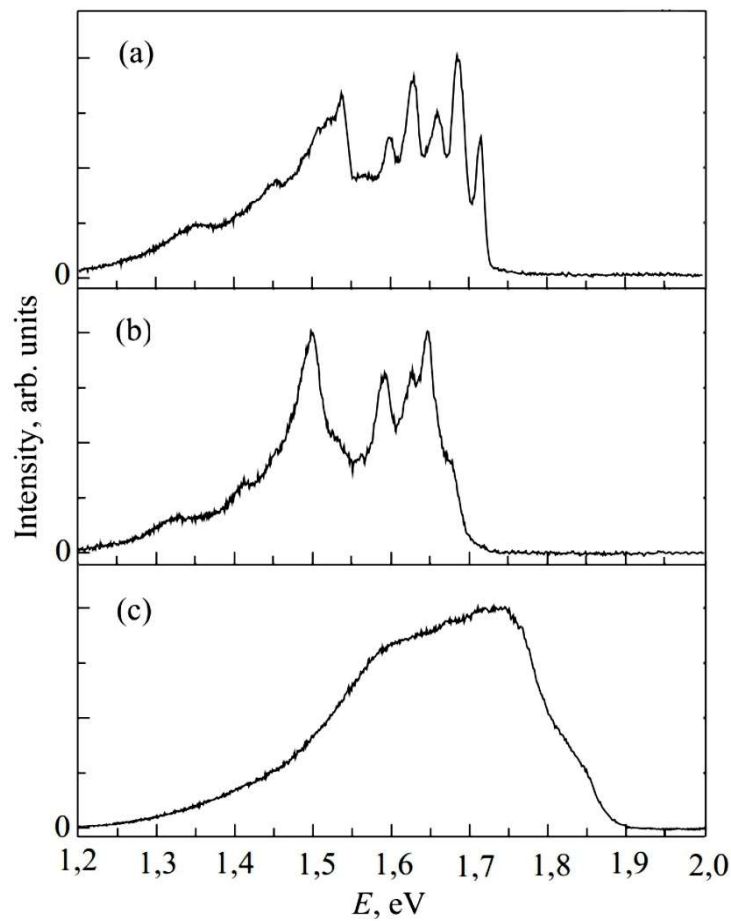

**Figure 1.** Photoluminescence spectra of $C_{60}$ single crystals at T=10 K, excitation with a He–Ne laser ($E_{exc}$ = 1.96 eV): pure fullerite (a); intercalated with hydrogen at 200 °C and a pressure of 30 atm (b); intercalated with hydrogen at 300 °C and a pressure of 30 atm (c). The spectra were recorded under identical conditions and were not corrected for the spectral sensitivity of the setup.

All the observed changes in the spectral characteristics clearly indicate the process of hydrogenation of fullerite molecules under the selected saturation conditions. The validity of this conclusion is confirmed by the fact that the initial form of the photoluminescence spectra of the samples is not restored after their annealing in a dynamic vacuum of $10^{-3}$ mm Hg at temperatures up to 500 °C. The hydrogenation reaction at such relatively low temperatures (300 ºC > T > 250 ºC) becomes possible, most likely, due to the fact that fullerite itself is a catalyst for the reaction and promotes the dissociation of $H_2$ molecules with the subsequent formation of hydrofullerene molecules $C_{60}H_x$. Depending on the selected chemical sorption temperature, the samples contain predominantly one of the stable forms of molecules $C_{60}H_{18}$, $C_{60}H_{24}$, $C_{60}H_{36}$, $C_{60}H_{44}$, $C_{60}H_{52}$, but most often it is a mixture of several components at once. The presence of molecules with different degrees of hydrogenation is accompanied by an increase in non-uniform stresses in the crystallites and, naturally, leads to a diffuse type of photoluminescence spectrum and its shift to the high-energy region.

Experiments on the study of the behaviour of the fullerite luminescence intensity with temperature changes have repeatedly shown their exceptional informativeness for understanding the features of the dynamics of transfer and relaxation of electronic excitations. Increased sensitivity of the temperature dependence of the radiation intensity to phase and orientational transitions allows us to analyse the effect of specially introduced impurities on the dynamics of the fullerite crystal lattice. As mentioned above, the coherent transfer of electronic excitation to the luminescence emission centres of $C_{60}$ is provided directly in the low-temperature phase of the orientational glass. Consequently, high radiation intensity in pure fullerite is observed when the fullerene molecules retain their favourable mutual orientation. Temperature disordering of molecules leads to a change in the course of the temperature dependence of the radiation intensity, and a rather sharp drop is observed at the glass transition point $T_g$. Figure 2 shows the temperature dependences of the integrated intensity of photoluminescence radiation $I$(T) for the discussed samples: pure fullerite $C_{60}$ and saturated with hydrogen at temperatures of 200 ºC and 300 ºC. It is evident that the orientational glass phase in pure $C_{60}$ is preserved up to 90 K (green curve) as the temperature increases, in contrast to fullerite saturated with hydrogen molecules $H_2$ at 200 ºC (red curve). For these samples, an increase in $T_g$ was observed, which is caused by the growing interaction in the impurity-matrix system. This is a consequence of the presence of more than one hydrogen molecule in the octahedral cavity, which is typical for extremely saturated solid solutions of $C_{60}+H_2$. Rotational reorientations of neighbouring molecules are difficult and require more energy in this case. Consequently, the orientation of neighbouring molecules, which is favourable for excitation transfer in the crystal, and hence the radiation intensity, remains stable up to higher temperatures. With an increase in the intercalation temperature of fullerite $C_{60}$ to 300 ºC in an atmosphere of molecular hydrogen, a change in the sorption mechanism occurs (adsorption crossover). Diffusion filling of intermolecular cavities of the crystal lattice (physisorption) is replaced by chemical interaction of $H_2$ with fullerene $C_{60}$ molecules (chemisorption), resulting in the formation of fullerite hydride $C_{60}H_X$. It was established for the first time that fullerite hydrogenation leads to a change in the course of the dependence of the integral radiation intensity $I(T)$ on temperature. In this case, it changes slightly in the temperature range of 20–230 K and has no peculiarities in the region of orientational glass formation.

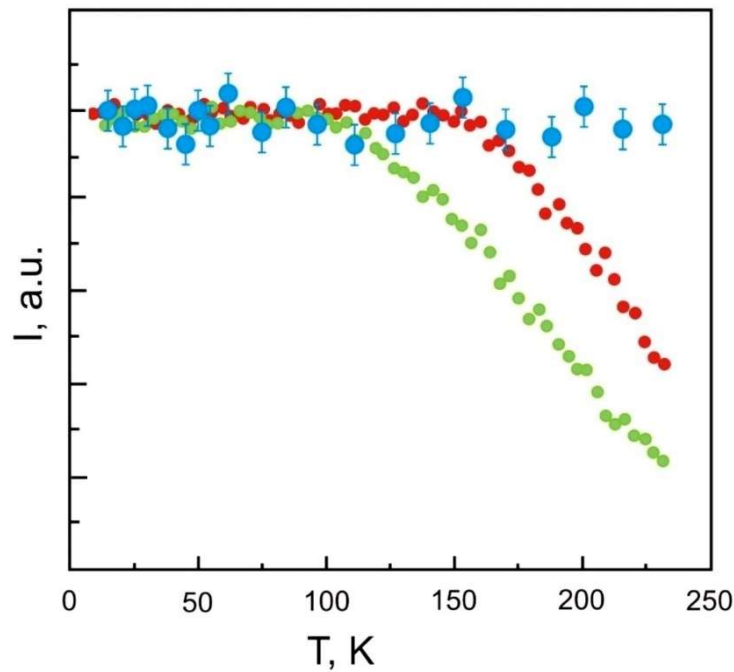

**Figure 2.** Normalized temperature dependences of the integrated radiation intensities of fullerite $C_{60}$: green - pure $C_{60}$; red - intercalated with $H_2$ at 200 ºC 30 atm., (physical sorption); blue - intercalated with $H_2$ at 300 ºC 30 atm., (chemical sorption).

In comparison with the extremely saturated solid solutions of $C_{60}+H_2$ obtained in the physical sorption mode, fullerite hydride $C_{60}H_X$ is a material with a higher hydrogen content. These molecules have lower symmetry and stronger intermolecular interaction compared to fullerene. As a result of the latter, as follows from the data of work [3], in hydrogenated fullerite up to 300 K there is no orientational phase transition.

### 3.2 $C_{60}+N_2$ system.

Figure 3 shows the low-temperature (30 K) photoluminescence spectra of pure $C_{60}$ and nitrogen-saturated polycrystalline samples (at 280 °C and 450 °C under a pressure of 30 atm), corrected for the spectral sensitivity of the setup and normalized to their integral intensity. The mobile

Frenkel exciton localized on "deep X-traps" makes the main contribution to the luminescence of the $C_{60}+N_2$ solution with a diffusion sorption mechanism. The intensity of such radiation depends on the concentration of intermolecular voids in the fullerite lattice filled with an intercalant. A noticeable increase in the total radiation intensity and significant inhomogeneous broadening of the spectral bands in Fig. 3(b) indicate a significant filling of the octahedral voids of the fullerite with $N_2$ molecules, at least at the penetration depth of the exciting laser radiation. In this case, X-ray structural data [4] show a relatively weak integral change in the lattice parameter throughout the volume of the polycrystalline sample. This indicates the obvious presence of a strong density gradient of the $N_2$ impurity from the surface into the depth of the $C_{60}$ crystals. In contrast to the above case of hydrogen physical sorption, there is no energy shift of the characteristic photoluminescence spectral bands.

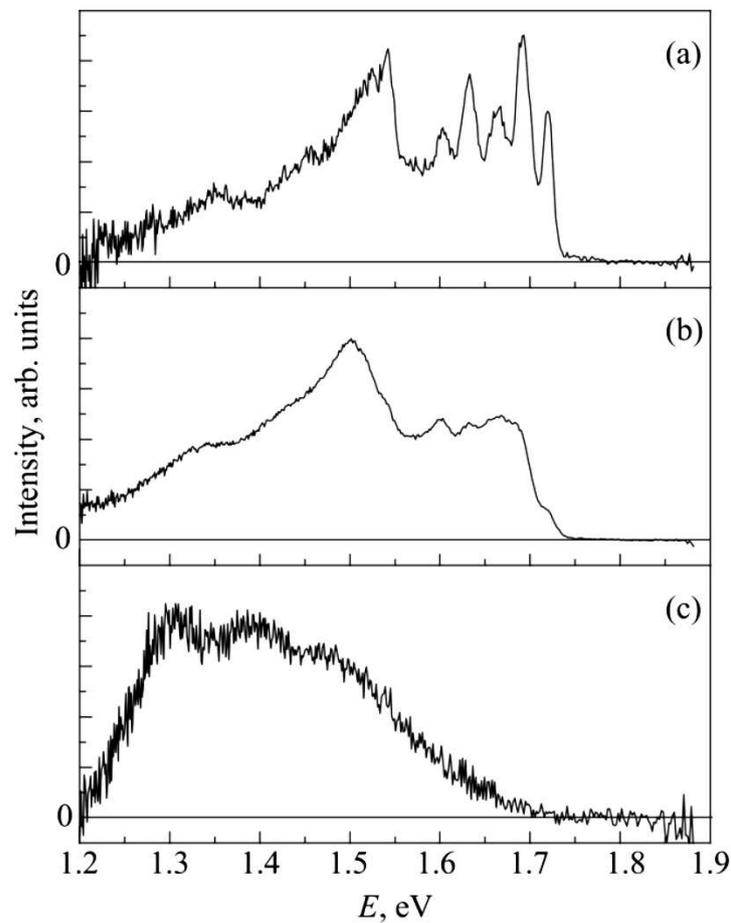

**Figure 3.** The photoluminescence spectra of a polycrystalline $C_{60}$ samples at 30 K: (a) pure fullerite; (b) $C_{60}$ saturated with molecular nitrogen at a temperature of 280 °C; (c) saturated with molecular nitrogen at a temperature of 450 °C. The spectra were normalized to their integral intensity and corrected for the spectral sensitivity of the setup.

This is a sign of weak interaction both in the impurity-matrix system and between neighboring $C_{60}$ fullerene molecules. When comparing the spectra of pure $C_{60}$ and fullerite extremely saturated with nitrogen at two different temperatures, Fig. 3(b) and Fig. 3(c), the abrupt nature of the change in photoluminescent properties with increasing intercalation temperature becomes obvious. And if for the sample saturated at a temperature of 280 °C only changes in the quantitative parameters of the radiation are observed, then for the second sample with a saturation temperature of 450 °C we observe a radical rearrangement of the spectrum with a strong shift to the low-energy region and a noticeable decrease in the radiation intensity. A joint analysis of these spectral-luminescent and structural [4] data (including vacuum degassing) leaves no doubt about the presence of a sorption crossover. An increase in the saturation temperature above 420 °C for the $C_{60}+N_2$ system leads to a change in the diffusion mechanism of sorption to a chemical interaction of the impurity with the matrix molecules and, as a consequence, to the formation of a new nitrogen-containing substance, and possibly a mixture of several different substances based on $C_{60}$. The primary analysis of the photoluminescence spectra showed the presence of bands characteristic of biazafullerite - $(C_{59}N)_2$.

To obtain more complete information on the electronic processes and excitation transfer features in the complex under study, the temperature dependences of the integral radiation intensities were recorded. Figure 4 shows such dependences for pure $C_{60}$ fullerite (green curve) intercalated with nitrogen under a pressure of 30 atm at 280 ºC (red curve) and saturated with nitrogen under a pressure of 30 atm at 450 ºC (blue curve). The study was carried out in the temperature range of 20–230 K in the modes of successive decrease and increase in temperature. Despite the fact that the rotational dynamics of $C_{60}$ molecules in impurity and pure crystals differs, the temperature dependence of the integral photoluminescence intensity always retains a certain character in this temperature range. As mentioned above, all previously studied two-component systems based on $C_{60}$, obtained by the diffusion sorption mechanism, are characterized by both a constant value of the quantum yield from low temperatures to the glass transition temperature and a subsequent decrease in intensity. A detailed analysis of the crystal lattice dynamics from the point of view of coherent transport of electronic excitations, as well as the mechanisms of the influence of impurities on it, are given in [5]. It should be immediately noted that the opposite shift of $T_g$ is observed here compared to the extremely saturated solutions of fullerite $C_{60}$ with hydrogen (red curves in Fig. 2 and Fig. 4). This clearly illustrates the above-discussed different influence of hydrogen and nitrogen impurities on the rotational dynamics of $C_{60}$ molecules in the crystal lattice. When octahedral cavities are filled with two molecules, hydrogen has an "inhibiting" effect on the surrounding fullerene molecules. Single-particle filling of the intermolecular space with $N_2$ impurity, on the contrary, facilitates mutual reorientations in the matrix nodes. And this, in the case of hydrogen, increases the glass transition temperature, and in the case of nitrogen, on the contrary, lowers it. As can be seen, the nature of the dependence for $C_{60}$ chemically saturated with nitrogen (blue curve) differs significantly from the results for pure fullerite and $C_{60}+N_2$ interstitial solutions. By the nature of the behavior, the curve for the chemical sorption mode can be conditionally divided into three sections starting from low temperatures: 25–50 K, where the intensity is stably low; 50–100 K, where a rapid increase in intensity is observed; 100–230 K, where a smooth decrease in the integral radiation intensity is observed.

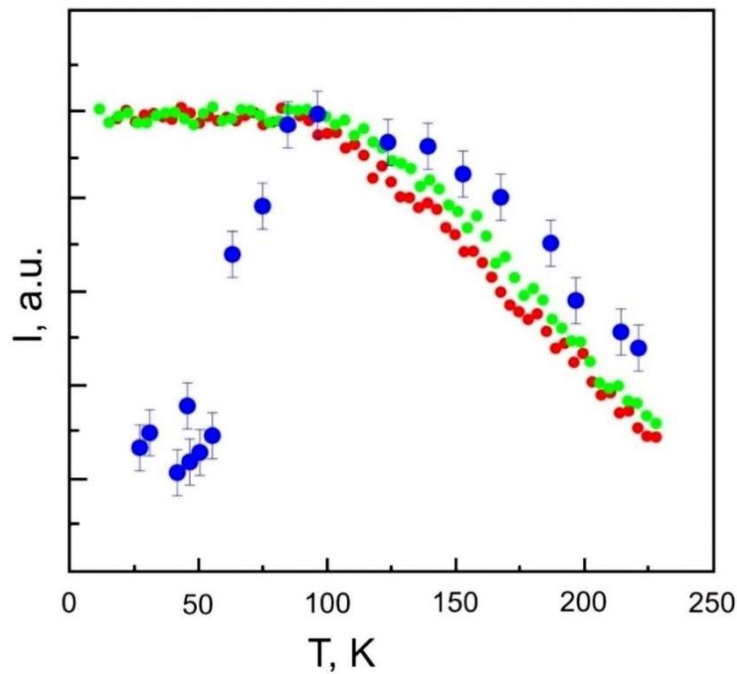

**Figure 4.** Temperature dependences of the integrated radiation intensities of $C_{60}$ fullerite, normalized to the corresponding values at T = 25 K: pure (green), intercalated $N_2$ at 280 ºC and a pressure of 30 atm (physical sorption) (red), intercalated $N_2$ at 450 ºC and a pressure of 30 atm (chemical sorption) (blue). All curves are measured in heating mode.

The course of this curve remains unchanged both in the heating and in cooling modes of the sample. Such a complicated behavior of the temperature dependence of the integral radiation intensity is clearly caused by the presence of additional mechanisms of electron excitation transport characteristic of a new nitrogen-containing substance based on fullerite $C_{60}$. Low photoluminescence intensity in the low-temperature region can be due to the emergence of complex molecular complexes during the nitriding of fullerene molecules, which play the role of quenching centers with a certain activation energy and a long lifetime. Apparently, such formations at low temperatures are most likely capable of capturing mobile Frenkel excitons with subsequent deactivation of excitation. Naturally, with increasing temperature, the energy contribution of the vibrational states of these centers increases and, consequently, the probability of releasing localized electronic excitation increases. This most likely determines the main mechanism for increasing the emission intensity in the region of 50 – 100 K, when excitons are already capable of leaving the potential well and migrating along the crystal, reaching the emission centers. With a further increase in temperature, a smooth decrease in the luminescence intensity is observed, the mechanism of which is probably similar to that for pure $C_{60}$ or interstitial solutions [5].

## 4. Conclusions

A combined analysis of the results of spectral-luminescent studies and studies of the temperature behavior of the photoluminescent properties of fullerite $C_{60}$ with impurities of molecular hydrogen and nitrogen confirms the presence of an adsorption crossover for both elements when changing the saturation temperature from the gas phase under a pressure of 30 atm.

For the systems $C_{60}+H_2$ and $C_{60}+N_2$, the diffusion saturation mechanism is replaced by chemical interaction in the region of 250 ºC and 420 ºC, respectively. The results of spectral and temperature studies of photoluminescence indicate the formation of new chemical compounds of fullerite with $H_2$ and $N_2$ at saturation temperatures above those indicated.

Comparison of the transformation of spectral-luminescent characteristics and their temperature behavior for the extremely saturated complexes $C_{60}+H_2$ and $C_{60}+N_2$ in the diffusion mode, relative to pure fullerite $C_{60}$, showed the opposite effect of hydrogen and nitrogen on the rotational subsystem of the crystal structure of the matrix and the efficiency of excitation transfer, respectively. If hydrogen complicates mutual reorientations of neighboring matrix molecules and increases $T_g$, then nitrogen, on the contrary, facilitates them and thereby lowers $T_g$.

For the first time, the temperature dependence of the luminescence intensity of hydrofullerite $C_{60}H_x$ was studied and its independence from temperature was recorded. Such behavior of the integral radiation intensity is explained by the absence of a transition to a glassy state in hydrofullerite, mainly due to the increased interaction of molecules in the lattice.

Analysis of the spectra of the $C_{60}+N_2$ system, for the case of nitrogen saturation in the chemisorption mode, showed the presence of biazafullerite $(C_{59}H)_2$ in the resulting multicomponent material. Significant differences in the behavior of the temperature dependence of the integrated radiation intensity of the new complex indicate the presence of additional mechanisms of electron excitation transport. The photoluminescence quenching at low temperatures found here may be associated with the emergence of new exciton capture centers with high efficiency of nonradiative deactivation of electron excitation.